\documentclass{cjaa}                   

\usepackage{graphicx}                   
\input{epsf.sty}                        
\input{psfig.sty}                       

\begin{document}
 \title{Effects of Magnetic Fields on Neutrino-dominated Accretion Model for
 Gamma-ray Bursts
}

   \volnopage{Vol.0 (200x) No.0, 000--000}      
   \setcounter{page}{1}          

   \author{Yi Xie
      \inst{}\mailto{}
   \and Chang-Yin Huang
   \inst{}
     \and Wei-Hua Lei
   \inst{}
   }
   \offprints{Yi Xie}                   

   \institute{Department of Physics, Huazhong University of Science and Technology, Wuhan
430074, China\\
\email{sourcexieyi@gmail.com}}
   \date{ }

   \abstract{
Many models of gamma-ray bursts suggest a common central engine: a
black hole of several solar masses accreting matter from a disk at
an accretion rate from 0.01 to 10 $M_\odot s^{-1}$. The inner
region of the disk is cooled by neutrino emission and large
amounts of its binding energy were liberated, which could trigger
the fireball. We improve the neutrino-dominated accreting flows by
considering the effects of the magnetic fields, and find that more
than half of the liberating energy can be extracted directly by
the large-scale magnetic fields on the disk. And it turns out that
the temperature of the disk is a bit lower than the
neutrino-dominated accreting flows without magnetic field.
Therefore, The outflows are magnetically-dominated rather than
neutrino dominated. In our model, neutrino mechanism can fuel some
GRBs (not the brightest ones), but cannot fuel X-ray flares.
However, the magnetic processes (both BZ and electromagnetic
luminosity from a disk) are viable mechanisms for most of GRBs and
the following X-ray flares.
   }
\authorrunning{Yi Xie, Chang-Yin Huang {\&} Wei-Hua Lei}            
   \titlerunning{Effects of Magnetic Fields on Neutrino-dominated Accretion Model }  

  \maketitle

%
%

\section{Introduction}           
\label{sect:intro}

\quad\quad Gamma-ray bursts (GRBs) are flashes of gamma-rays
occurring at cosmological distances, being the most powerful
explosions since the Big Bang. They are generally divided into two
classes (Kouveliotou et al.1993): short-duration ($T_{90} <2s$)
hard-spectrum GRBs (SGRBs) and long-duration ($T_{90} >2s$)
soft-spectrum GRBs (LGRBs), which have different progenitors.
LGRBs root in core collapses of massive, rapidly rotating stars
(Woosley 1993, Paczynski 1998, Hjorth et al 2003, Stanek et al
2003), and supernovae have been observed coincidently in some
LGRBs (Galama et al 1998; Stanek et al 2003; Hjorth 2003). In
contrast to LGRBs, SGRBs may arise from coalescence of neutron
stars or black hole binary systems due to damping of gravitational
radiation (e.g. Eichler et al. 1989, Narayan, Paczynski \& Piran
1992, Fryer \& Woosley 1998), and they are probably associated
with elliptical galaxies (Gehrels et al. 2005; Bloom et al 2006;
Barthelmy et al. 2005; Berger et al 2005). It is believed that the
two processes give rise to a black hole of several solar masses
with a magnetized disk or a torus around it (Meszaros \& Rees
1997b). And many central engine models of GRBs based on this
scenario (exception models, for instance, magnetizedrotating
neutron stars, see e.g. Usov 1992).

Some authors have studied the accretion model for GRBs by assuming
steady-state accretion (e.g. Papham, Woosley \& Fryer 1999,
hereafter PWF; Narayan, Piran \& Kumar 2001, hereafter NPK; Di
Matteo, Perna \& Narayan 2002, hereafter DPN). Their studies show
that at the extremely high accretion rate (0.01 to 10 $M_\odot
s^{-1}$) needed to power GRBs, the disk cannot be cooled
efficiently as the gas photon opacities are very high, and a large
fraction of its energy is advection dominated. However, inner
region of the disk becomes hot and dense enough to cool via
neutrino emission, and this accretion mode is referred to as
neutrino-dominated accretion flows (NDAFs). The neutrinos can
liberate large amounts of binding energy via the
$\nu\bar{\nu}\rightarrow e^+e^-$ processes in regions of low
baryon density and then trigger the fireball.

However, the model with "neutrino-driven outflow" alone cannot be
a candidate of some GRBs central engine. For instance, numerical
simulations by Shibata et al. (2006) suggest that the collapse of
hypermassive neutron-star triggered by gravitational wave cannot
be a candidate for the central engine of SGRBs, however, it
becomes powerful enough to produce the fireball after taking the
magnetic braking and MRI into account. On the other hand,
researches show that the magnetic fields can be magnified up to
$10^{15}\sim10^{16} G$ by virtue of magnetorotational instability
(MRI, Balbus \& Hawley 1991) or dynamo processes (Pudritz \&
Fahlman 1982 and references therein) in the inner region of the
disk. So, the existence of strong magnetic fields should be
considered. Both PWF and DPN compared the luminosity of neutrino
emission and Poynting flux, and indicated that MHD processes are
viable mechanisms for powering GRBs, but they did not include
magnetic fields in their disk conditions.

In this paper, we intend to improve the NDAF model by considering
the effects of magnetic braking and magnetic viscosity. The
equation of angular momentum of a standard disk is replaced by the
equation of a magnetized disk in which the viscosity caused by
magnetic braking and magnetic viscosity only. Meanwhile, we deduce
the rotational energy extracted by large-scale fields in the disk
from the thermal energy produced by viscous dissipation, and the
magnetic pressure is considered in equation of state else. It
turns out that the inner region of the disk is magnetically
dominated. Magnetized accretion models within the GRB context have
been also discussed in several other papers in which detailed
numerical simulations are performed. For example, Proga et al.
(2003) studied a MHD collapsar accretion model and suggested that
MHD effects alone can launch a GRB jet, which is Poynting flux
dominated. Mizuno et al. (2004a; 2004b) drew similar conclusions
using a GR-MHD code. These results agree with our conclusion in
this paper.

This paper is organized as follows. In Section 2 we outline the
theory of a magnetized accretion disk. In Section 3 we introduce
the basic assumptions and equations of our model. In Section 4 we
show the numerical results of our model, and finally, In Section 5
we summarize main results of our model and some defects are
discussed.

\section{DESCRIPTION OF A MAGNETIZED ACCRETION DISK }
\label{sect:Des}

\quad\quad It is widely known that the magnetic fields on the disk
can greatly affect angular momentum transfer and hence the
accretion rate via a variety of modes. In this paper we only
consider two basic mechanisms: the first one is magnetic
viscosity, the weak magnetic fields creates MRI in which
turbulence dominates the angular momentum transfer (Balbus \&
Hawley 1991). The second is the magnetic braking, the large scale
magnetic field extracts rotational energy of disk due to the shear
force of differential rotation (Blandford 1976; Blandford \& Payne
1982; Livio et al 1999). We assume that the accretion process is
governed by these two mechanisms completely. The main points of
the magnetized accretion disk model (in which the magnetic braking
and magnetic viscosity are considered only for angular momentum
transfer) given by Lee et al. (2000) are outlined as follow.

According to Torkelsson et al. (1996) the magnetic viscosity
$\nu^{mag}$ is defined as

\begin{equation}
\label{eq1} B_\phi B_r/ 4\pi = -\nu^{mag}\left( r
d\Omega_{disk}/dr
\right)\rho ,\\
\end{equation}
where $B_\phi$ and $B_r$ are respectively the azimuthal and radial
components, $\Omega_{disk}$ is angular velocity and $\rho$ is the
density of the disk matter. The magnetic viscosity $\nu^{mag}$ can
be parameterized as (Shakura \& Sunyaev 1973; Pringle 1981)

\begin{equation}
\label{eq2} \nu^{mag} = \alpha^{mag} c_s H ,\\
\end{equation}
where $c_s$ is the sound velocity of the disk ($c_s = \left(
P_{disk} / \rho \right)^{1/2}$ ,  in which $P_{disk}$ is the disk
pressure) and $H$ is the half-thickness of the disk. Invoking
hydrostatic equilibrium perpendicular to the disk plane, we have
$H = c_s/\Omega_{disk}$ and

\begin{equation}
\label{eq3} \nu^{mag} = \alpha^{mag} c_s^2/\Omega_{disk} .\\
\end{equation}

For a Keplerian orbit we have $\Omega_{disk}\sim\Omega_{K} =
\left( G M / r^3\right)^{1/2}$, and Eq. $\left( 1 \right)$ can be
written as

\begin{equation}
\label{eq4} B_\phi B_r/ 4\pi =  \frac {3}{2}  \alpha^{mag} P_{disk} .\\
\end{equation}

The accretion rate is generally determined by magnetic braking for
$H\ll r$ (Lee, Wijers \& Brown 2000), the angular momentum balance
equation can be written as

\begin{equation}
\label{eq5} \dot{M} = 2 r B_\phi B_Z/\Omega_{disk} .\\
\end{equation}

The axisymmetric solution (Blandford 1976) is

\begin{equation}
\label{eq6} B_\phi = 2 r \Omega_{disk} B_Z/ c .\\
\end{equation}

A roughly steady state will be reached when the grown rate of
$B_\phi$ generated by differential rotation from radial field
equals to its loss rate by buoyancy, then the magnitude of
$B_\phi$ can be estimated as (Katz 1997)

\begin{equation}
\label{eq7} B_\phi \approx  \left[\frac {3}{2} B_r
\Omega_{disk}H\right]^{1/2}\left( 4\pi\rho\right)^{1/4} .\\
\end{equation}

By using Eqs. (4), (6) and (7) we have

\begin{equation}
\label{eq8} B_Z = \frac {c}{2} \left( \frac {\pi r P_{disk}}
{ G M}\right)^{1/2}\left(9 \alpha^{mag} \right)^{1/3} ,\\
\end{equation}
and
\begin{equation}
\label{eq9} B_\phi = \left( \pi P_{disk}\right)^{1/2}\left
(9 \alpha^{mag} \right)^{1/3} .\\
\end{equation}

The vertical component and azimuthal component of field can be
estimated by using Eqs. (8) and (9), and for a given
$\alpha^{mag}$, only depending on the gas pressure.

Combining Eqs. (6) and (8) with Eq. (5) we have (see also Lee,
Wijers \& Brown 2000)

\begin{equation}
\label{eq10} \dot{M} = 4 r^2 B_Z^2 / c .\\
\end{equation}

Since $\dot{M}$ is independent of $r$ in steady-state accretion,
we can infer that $B_Z \propto 1/r $.

\section{BASIC ASSUMPTIONS AND EQUATIONS OF MODEL}
\label{sect:Bas}

\quad\quad The basic physical conditions in disk models for GRBs
can be derived by virtue of steady-state conditions (PWF, NPK,
DPN). Base on these studies, we consider the effects of magnetic
field in inner regions of disks in the frame of hydrodynamics. The
basic equations consist of equation of state and the conservation
equations of energy and angular momentum in a magnetized accretion
disk, which are described as follows.

In equation of state we include the contributions from radiation
pressure, gas pressure, degeneracy pressure and magnetic pressure,

\begin{equation}
\label{eq11} P = \frac{11}{12}aT^4 + \frac{\rho k T}{ m_p} +
 \frac{2 \pi h c }{3} \left( \frac{3}{8 \pi m_p} \right)^{4/3}
\left( \frac{\rho}{\mu_e}
\right)^{4/3} + \frac{B^2}{8 \pi} ,\\
\end{equation}
where $a$ is the radiation constant, $T$ is the disk temperature,
and the factor $\frac{11}{12}$ includes the contribution of
relativistic electron-positron pairs. In degeneracy term, $\mu_e$
is the mass per electron, and it is taken as $2$ by assuming equal
number of protons and neutrons. For the magnetic pressure we only
consider poloidal component in calculation, provided that it is
not much less than the toroidal component.

The conservation of mass is written by NPK and DPN as follows,

\begin{equation}
\label{eq12} \dot{M} = 4 \pi r \upsilon_r \rho H  \approx 6 \pi \rho \nu H ,\\
\end{equation}
where $\upsilon_r$ is the radial velocity and $\upsilon_r = 3\nu /
2 r$. Different from NPK and DPN, we replace the Eq. (12) by Eq.
(10), which includes the effects of magnetic braking and magnetic
viscosity.

In energy equation, the viscous heating equals neutrino radiative
loss plus advective loss and the fraction of rotational energy
extracted by large-scale magnetic fields

\begin{equation}
\label{eq13} \frac{3 G M \dot{M}}{8 \pi r^3} = \left( q_{\nu\bar
{\nu}}^- + q_{eN}^- \right)H + q_{adv} + Q_B^- ,\\
\end{equation}
in which $q_{\nu\bar {\nu}}^-$ is cooling via pair annihilation
and we take it as the approximation of Itoh et al. (1989; 1990):
$q_{\nu\bar {\nu}}^- \simeq 5\times 10^{33} T_{11}^9 ergs cm^{-3}
s^{-1}$ (in which $ X_{n} = X/10^{n}$ is used). The $q_{eN}^-$
represents the cooling via pair capture on nuclei, and can be
estimated as $q_{eN}^- \simeq 9 \times 10^{33} \rho_{10} T_{11}^6
ergs cm^{-3} s^{-1}$. And $q_{adv}$ is the advective cooling rate,
 we approximate it by (see eg., Narayan \& Yi 1994; Abramowicz
et al.1995)

\begin{equation}
\label{eq14} q_{adv} = \Sigma\nu T \frac{ds}{dr} \simeq  \xi \nu
\frac{H}{r} T \left( \frac{11}{3}aT^3  + \frac{3}{2}\frac
{\rho k T}{ m_p}\frac{1 + X_{nuc}}{4}\right) ,\\
\end{equation}
in which $s$ is specific entropy, $X_{nuc}$ is the mass fraction
of free nucleons, $\xi \propto -dlns/dlnr$ is assumed to be equal
to $1$ as in DPN. And finally, $Q_B^-$ represents the energy
extracted by magnetic field (see, Lee, Wijers \& Brown 2000)

\begin{equation}
\label{eq15} Q_B^- = \frac{dP^{mag}}{dS} = \frac{B_Z^2 r^2}{\pi
c}\left( \frac{G M}{r^3}\right) = \frac{G M \dot{M}}{4 \pi r^3} ,\\
\end{equation}
where $dS = 2 \pi r dr$. Comparing Eq.(15) with Eq.(13), we find
that two thirds of energy of viscous heating was substituted by
the field extracting, therefore, it is magnetically dominated in
the inner regions of disks.

Eqs. (8), (10), (11) and (13) contain four independent unknowns
$P$, $\rho$, $T$ and $B_Z$ as functions of $r$ and compose a
complete set of equations which can be numerically solved with
given $M$, $\alpha$ (for simplicity, we omit the superscript
'mag') and $\dot{M}$. In the following calculations we fix $M = 3
M_\odot$ (the corresponding Schwarzschild radius $R_s$ is $2
GM/c^2 = 8.85 \times 10^5 cm )$, $\alpha = 0.1$ .

\section{NUMERICAL RESULTS}
\label{sect:Num}
\subsection{Gas Profiles}
\label{sect:Gas} \quad\quad We show the numerical solutions of the
full equations in this section, and the software of "Mathematica"
is used for the numerical algorithm of Newton iteration method.
The pressure components profiles are shown in Figure $1$, and the
solutions for three values of the accretion rate $\dot{m} = 0.1$,
$1$, and $10$ ($\dot{m}$ is defined as $\dot{m} = \dot{M} /
M_\odot s^{-1}$) are show in (a), (b) and (c) respectively. The
gas pressure, degeneracy pressure, radiation pressure, and
magnetic pressure are shown by the solid line, dotted line, dashed
line, and long-dashed line, respectively. From Figure $1$, we
obtain the following results:

\begin{figure}
\vspace{1.0cm}
\begin{center}
\includegraphics[width=4.3cm]{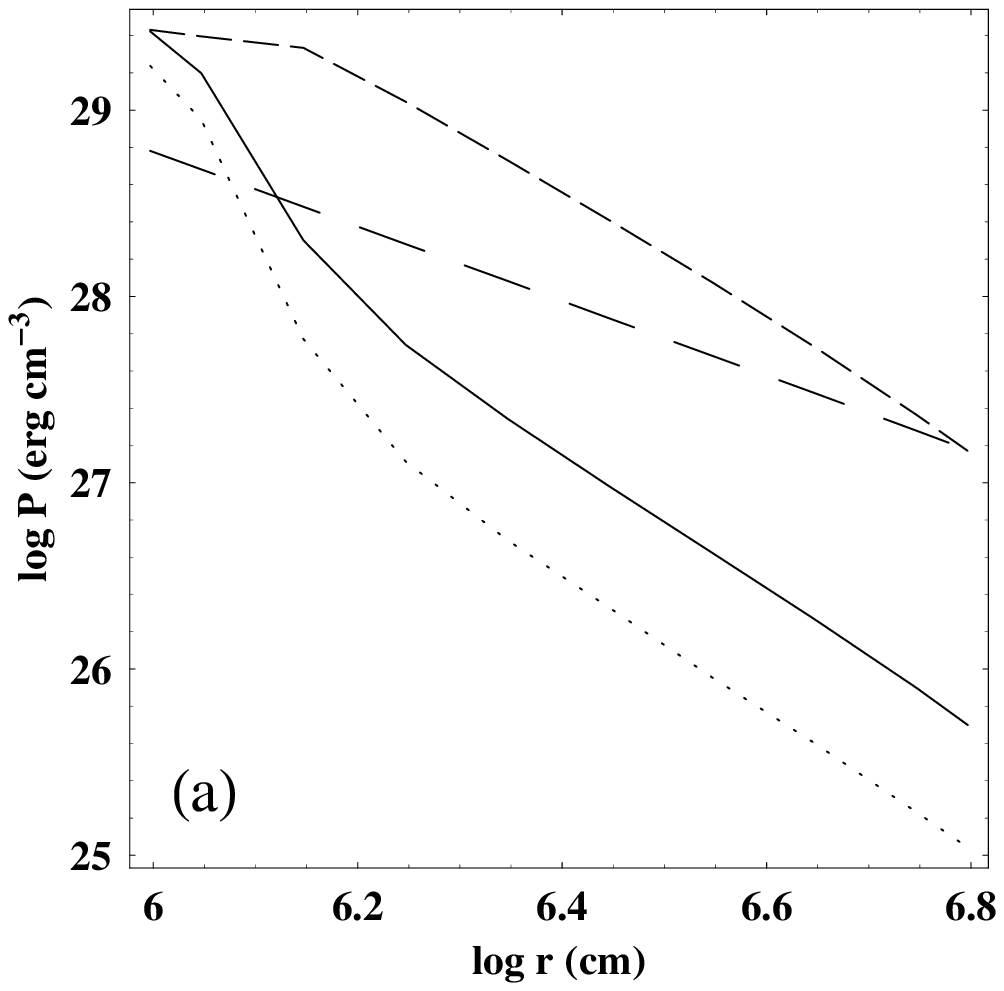}
 \includegraphics[width=4.3cm]{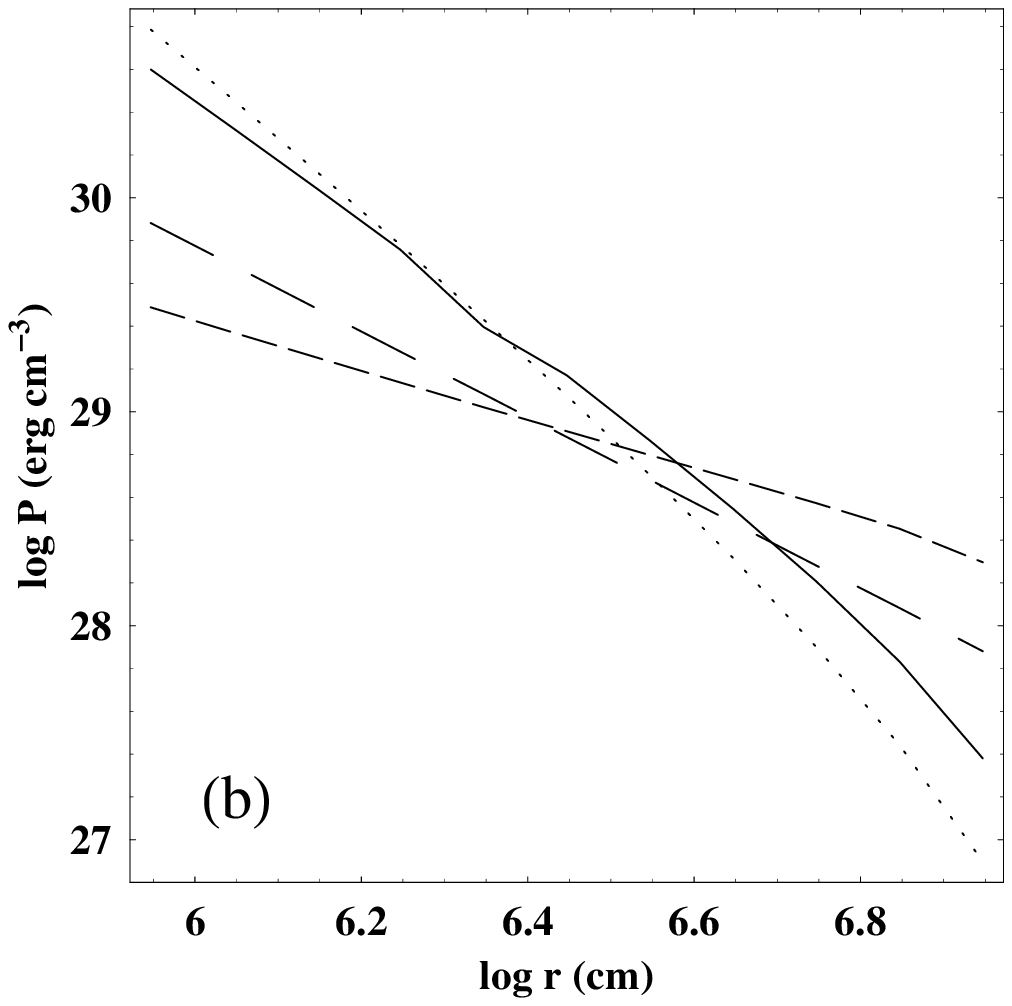}
 \includegraphics[width=4.3cm]{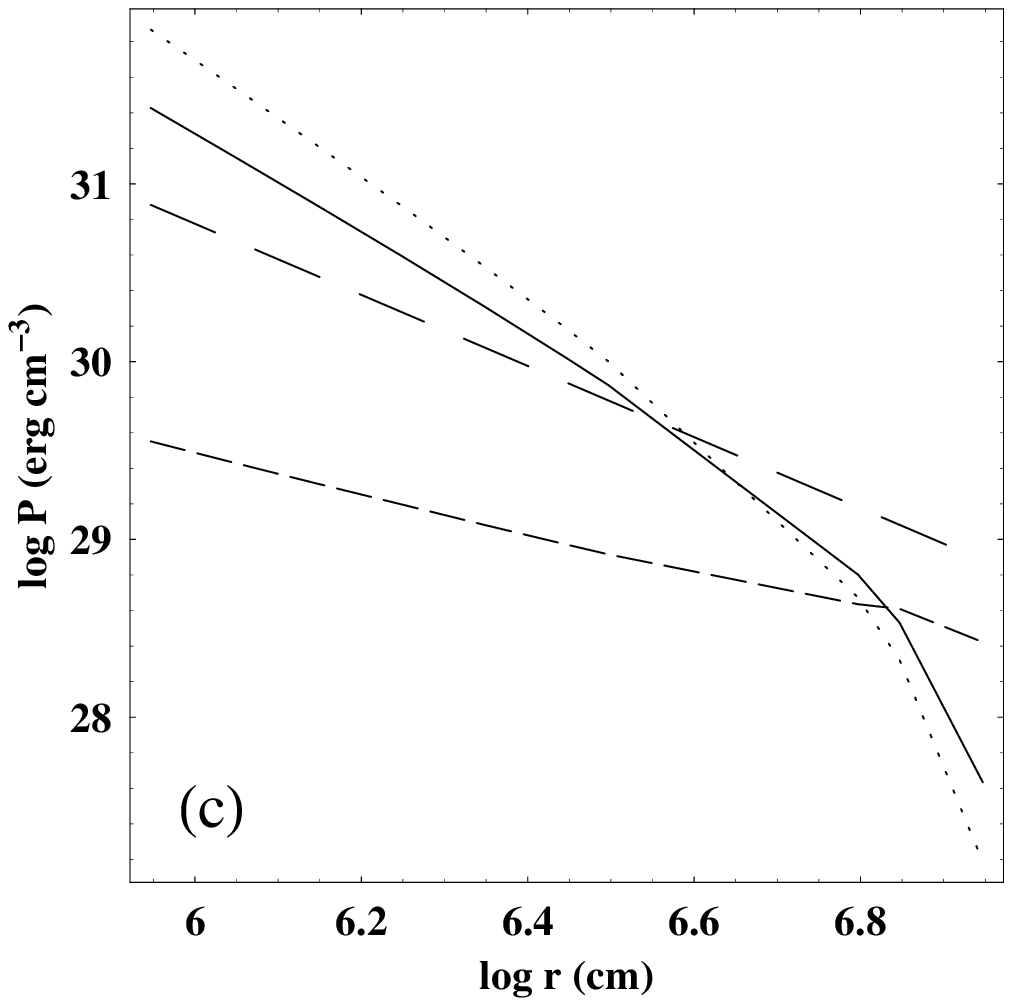}
 \caption{Pressure components profiles for three values of the
 accretion rate: (a) $\dot{m} = 0.1$, (b) $\dot{m} = 1$, and
 (c) $\dot{m} = 10$. The gas pressure is shown
  by the solid line, degeneracy pressure by dotted line,
  radiation pressure by dashed line, and magnetic pressure by
   long-dashed line.}
\end{center}
\end{figure}

(i) From (a) we can see that, the flows is radiation pressure
dominated in the region of $1 R_S \sim 10 R_S$  and may be
thermally unstable (see 4.3, stability analysis). It is thermally
stable in the same region in DPN as it always dominates by gas
pressure.

(ii) From (b) and (c) we can see that, the magnetic pressure
component is more important at large radii and even overwhelms the
gas pressure and degeneracy pressure. So our model is valid only
in a narrow region because of the restriction of Eq. (12).

Temperature and density profiles calculated from our model are
shown in Figure $2$. We show our solutions for three values of the
accretion rate, $\dot{m} = 0.1$, $1$, and $10$ (long dashed,
solid, and short-dashed lines, respectively). Comparing with DPN
(see DPN, Fig.1), we find that the temperature of disk is a bit
lower than NDAFs without considering the effects of magnetic
fields, and the density drops much more rapidly with the radius.

\begin{figure}
\vspace{1.0cm}
\begin{center}
\includegraphics[width=5cm]{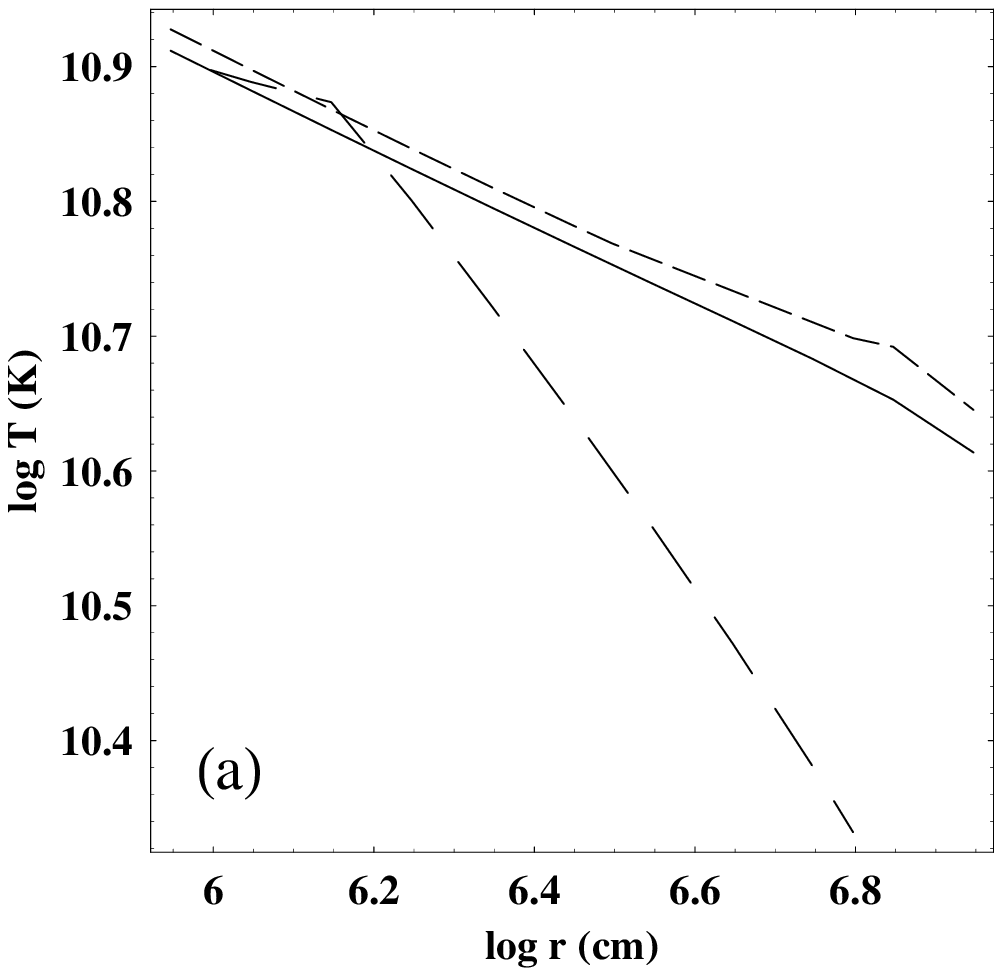}
 \includegraphics[width=5cm]{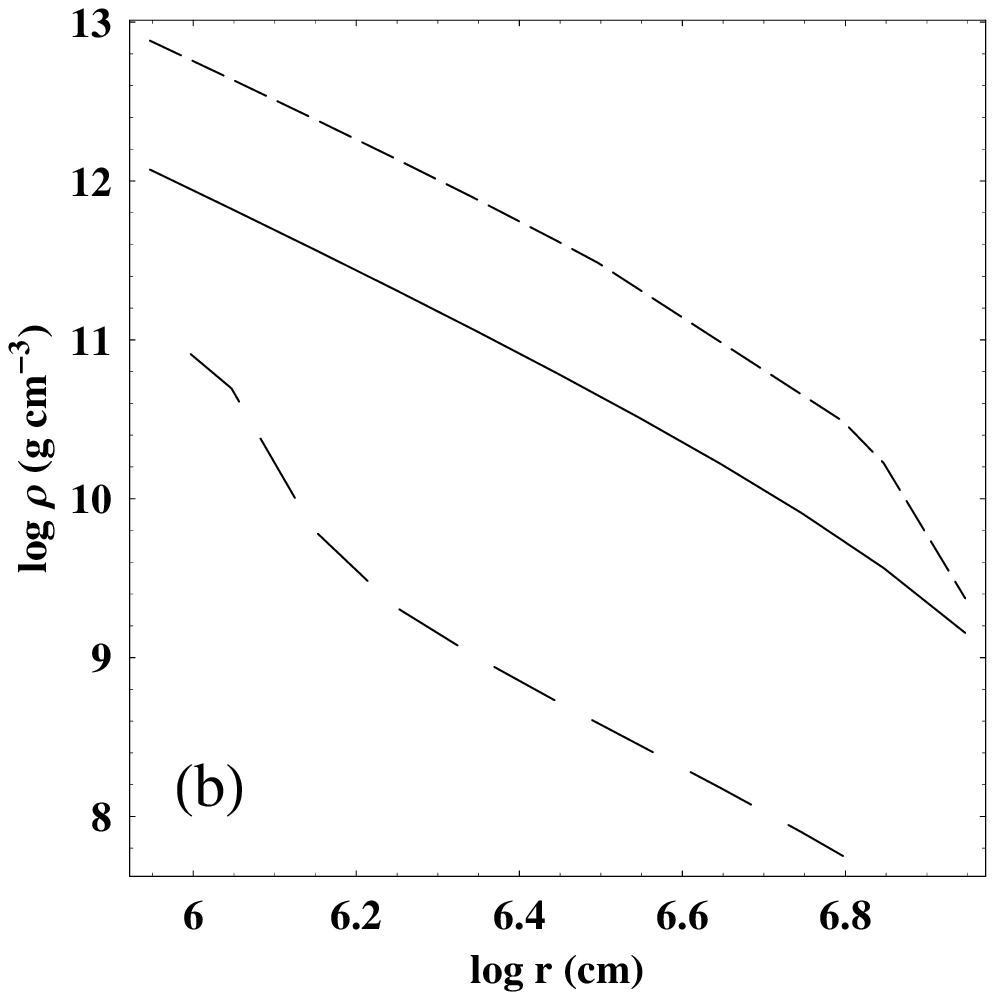}
 \caption{Temperature and density profiles in (a) and (b) respectively.
The profiles are shown for three values of the accretion rate:
$\dot{m} = 0.1$ (long dashed lines), $\dot{m} = 1$ (solid lines),
and $\dot{m} = 10$ (short dashed lines).}
\end{center}
\end{figure}

\subsection{BZ Luminosity, Electromagnetic Luminosity from Disk,
and Neutrino Luminosity} \label{sect:BZ} \quad\quad It is a common
assumption that the magnetic fields will rise up to some fraction,
which for instance in DPN, $10\%$ of its equipartition value $B^2
/ 8 \pi \sim \rho c_s^2$. For $0.1 < \dot{m} <10 $, the typical
values of $\rho c_s^2$ are $10^{30} \sim 10^{32} ergs cm^{-3}$,
implying a field strength of $10^{15} \sim 10^{16} G$. The BZ jet
luminosity is then

\begin{equation}
\label{eq16} L_{BZ} = \frac{B_H^2}{4 \pi}\pi c a^2 R_S^2  \simeq
10^{52} a^2 \left( \frac{B_H}{10^{16} G} \right)^2 \left(
\frac{M}{3 M_\odot} \right)^2 ergs cm^{-3}
 ,\\
\end{equation}
in which $a$ is the dimensionless black hole spin parameter, $B_H$
is the magnetic field on the horizon .

The electromagnetic power output from a disk is equal to the power
of the disk magnetic braking can be calculated as (Livio et al.
1999; Lee et al. 2000)

\begin{equation}
\label{eq17} L_d = \frac {B_z^2}{4 \pi} \pi r^2 \left( \frac {r
\Omega_{disk}}{c}\right) c \approx a^{-2} \left( \frac {B_Z}{B_H}
\right)^2 \left( \frac {r}{R_S} \right)^{3/2} L_{BZ} .\\
\end{equation}
Consistent with previous work (Merloni \& Fabian 2002; DPN), we
take approximately,

\begin{equation}
\label{eq18} B_Z \sim \left( H / r \right) B_H .\\
\end{equation}
It is easy to get the strength of poloidal field in the disk and
the field on the black hole horizon for a given $\dot {m}$ by
using Eqs. (10) and (18) in our model, without the assumption of
equipartition value discussed above. And then, the BZ jet
luminosity and the electromagnetic power from a disk can be
calculated from Eqs. (16) and (17). The neutrino luminosity is
given by $L_\nu = \int_{r_{min}}^{r_{min}} 2 \pi q_\nu^- r dr$, in
which $q_\nu^- = \left( q_{\nu\bar {\nu}}^- + q_{eN}^- \right)H$,
$r_{min} = 1 R_S$ (for an extreme Kerr black hole), and $r_{min}=
10 R_S $. We estimate the luminosity due to $\nu \bar {\nu}$
annihilation along z-axis above the disk to be the Eq. (21) in
DPN. In Figure 3 we show the curves of BZ luminosity $L_{BZ}$
(solid lines), electromagnetic luminosity from a disk $L_d$ (
long-dashed line), and neutrino annihilation luminosity $L_{\nu
\bar{\nu}}$ (short-dashed line) versus dimensionless accretion
rate. From Figure 3, we obtain the following results:

\begin{figure}
\vspace{1.0cm}
\begin{center}
\includegraphics[width=8cm]{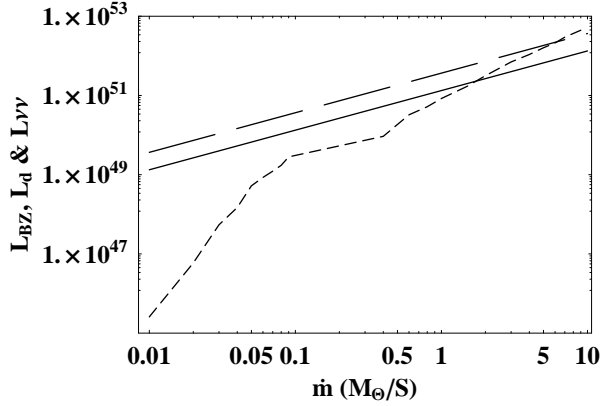}
 \caption{The solid lines represent the BZ luminosity, and the
 long-dashed line represent the electromagnetic luminosity from
 a disk and short-dashed line represents neutrino annihilation
 luminosity as a function of accretion rate and value of $H/r=0.2$.}
\end{center}
\end{figure}

(i) $L_d$ is larger than $L_{BZ}$ for $H/r=0.2$ (solution of this
model), however, both of them are viable mechanisms for central
engines of GRBs , and can also fuel the observed X-ray flares in
which case the accretion rate of $\dot{m} = 0.01$ is needed.

(ii) The $L_{\nu \bar{\nu}}$ is around $10^{51} ergs s^{-1}$ at
$\dot{m} = 1$, which is sufficient to power some GRBs, and at the
accretion rate of $\dot{m} = 0.01$, it fails to fuel the X-ray
flares.

In conclusion, neutrino mechanism can fuel some GRBs (not the
brightest ones), but cannot fuel X-ray flares. However, the
magnetic processes (both BZ and electromagnetic luminosity from a
disk) are viable mechanisms for most of GRBs and the following
X-ray flares (this agrees well with discussions of Fan et al.
(2005)).

\subsection{Stability}
\label{sect:Sta} \quad\quad Both NPK and DPN discussed the
stability properties of their solutions. Since our model considers
the effects of magnetic fields and differs considerably with
theirs, it is interesting to examine whether the solution is
stable.

The general condition for thermal stability is given by (Piran
1978)

\begin{equation}
\label{eq20} \left( \frac{d\ln Q^+}{d \ln T} \right)_{|\Sigma}
 < \left( \frac{d \ln Q^-}{d \ln T} \right)_{|\Sigma} ,\\
\end{equation}
in which $Q^\pm$ are the integrated (over the height of the disk)
heating (+) and cooling (-) rates. The cooling rate $Q^- = q_\nu^-
+ q_{adv} + Q_B^-$. We show the two curves of $Q^-$ and $Q^+$ as a
function of gas temperature in Figure 4. The radius is fixed at $r
= 5 R_S$, while the surface density is taken to be $\Sigma =
10^{16} g cm^{-2}$. From Figure 4 we can see that, the flow is
unstable while the temperature $T$ is lower than $5\times 10^{10}
K$, because the magnetic fields extract rotational energy from
disk is independent with temperature. When $T > 5\times 10^{10}
K$, it turns to stable because $q_{eN}^- \propto T^6$ becomes
relatively significant with the temperature increasing. When the
disk temperature crosses the critical point of the instability
curves, the thermal energy would be released suddenly in a thermal
time scales. It is possibly an explanation for the variability
time scales of tens of msec in the light curves of the GRBs, and
we will give the details in another paper.

\begin{figure}
\vspace{1.0cm}
\begin{center}
\includegraphics[width=6cm]{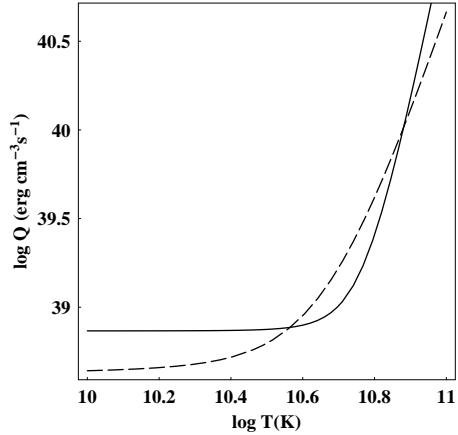}
 \caption{Thermal stability analysis. Dashed line shows $Q^+$  and
 solid line shows $Q^-$  vs. $T$ for $r
= 5 R_S$ curves, and for $\Sigma = 10^{16} g cm^{-2}$. The
solution is unstable while $T < 5\times 10^{10} K$, and becomes
stable when $T > 5\times 10^{10} K$. }
\end{center}
\end{figure}

Following NPK and DPN we use the condition for viscous stability

\begin{equation}
\label{eq21} \frac{d\dot{M}}{d \Sigma} > 0 .\\
\end{equation}
In our model, we have $\dot{M} \propto \Sigma$ for the case of gas
pressure dominated, $\dot{M} \propto \Sigma^{8/7}$ for degeneracy
pressure case, $\dot{M} \propto \Sigma^{2}$ for radiation pressure
case, and we have $\dot{M} \propto B^{2}\propto P$ for the
magnetic dominated and the surface density $\Sigma\propto
 P^{1/2}\rho^{1/2}$, meanwhile, considering the magnetic fields
decreases as $r^{-1}$, we have approximately $\rho \propto
 r^{-3}$, and then $\dot{M} \propto \Sigma$. All of these
cases are clearly viscously stable.

Finally, we also consider the gravitational instability. The
accretion flow will become gravitational unstable if the Toomre
parameter $Q_T$ is less than $1$, for a Keplerian orbit, which is
given by (Toomre 1964)

\begin{equation}
\label{eq22} Q_T = \frac{c_s \kappa}{\pi G \Sigma} = \frac{\Omega_K^2}{\pi G \rho} .\\
\end{equation}
We have checked that $Q_T\gg 1$  hence the flow is gravitationally
stable in the inner region of the disk. Nevertheless, at large
radii the Toomre parameter could be less than 1 as argued by Perna
et al. ( 2006 ). Actually that was another model for X-ray flares.

\section{CONCLUTION AND DISCUSSION}
\label{sect:Con} \quad\quad In this paper we modify the NDAFs
model as a central engines for GRBs by considering the effects of
magnetic braking and magnetic viscosity in the frame of Newtonian
dynamics. We found that two thirds of the liberating energy was
extracted directly by large-scale magnetic fields on the disk and
the temperature of a disk is a bit lower than the NDAFs without
magnetic fields. Furthermore, the density of the disk drops faster
than NDAFs along the radius. Therefore, the inner region of the
flow is magnetically dominated rather than neutrino dominated.
However, the neutrino mechanism can still fuel some GRBs (not the
brightest ones), but cannot fuel X-ray flares. However, the
magnetic processes (both BZ and electromagnetic luminosity from a
disk) are viable mechanisms for most of GRBs and the following
X-ray flares.

Our model is formulated invoking Newtonian potential, ignoring the
effects of general relativity which may be important in some
aspects (Gu et al. 2006) and neutrino opacity. Specially, the main
simplification of the analytic approach (this and other analytical
works) is the requirement of a steady state, which in reality
(e.g. numerical simulations) is not necessarily justified. An
example is that with magnetic fields, both numerical simulations
(Proga $\&$ Begelman, 2003) and some analytical arguments (Proga
$\&$ Zhang, 2006) suggest that the accretion flow may not always
in a steady state. Rather, magnetic fields accumulated near the
black hole can form a magnetic barrier that temporarily blocks the
accretion flow. This makes some dormant epochs at the central
engine. The breaking of the barrier would lead to restarting the
central engine, which is required to explain the recent Swift
observations of X-ray flares (for a review of Swift results and in
particular X-ray flares and their interpretations, see Zhang,
2007). In this paper, the poloidal component of magnetic fields is
$B_Z \propto 1/r$, which implies that the magnetic pressure drops
much slower than the other components and the calculations
indicate that the magnetic fields pressure could be dominant at
larger radii. In fact, such over-pressure magnetic fields are the
agent to form the magnetic barrier as reported by Proga $\&$
Zhang, which is needed to interpret the observed X-ray flares. The
unsteady state accretion model and the case of over-pressure
magnetic fields will be studied in our future work.

\begin{acknowledgements}
We would like to thank the referee, whose comments led to a
significant improvement of this work. This work is supported by
the National Natural Science Foundation of China under grants
10573006.
\end{acknowledgements}

\end{document}